\documentclass[10pt,conference]{IEEEtran}

\IEEEoverridecommandlockouts
\usepackage{cite}
\usepackage{amsmath,amssymb,amsfonts}
\usepackage{graphicx}
\usepackage{textcomp}

\usepackage{scalerel}
\usepackage{tikz}
\usetikzlibrary{svg.path}

\definecolor{orcidlogocol}{HTML}{A6CE39}
\tikzset{
  orcidlogo/.pic={
    \fill[orcidlogocol] svg{M256,128c0,70.7-57.3,128-128,128C57.3,256,0,198.7,0,128C0,57.3,57.3,0,128,0C198.7,0,256,57.3,256,128z};
    \fill[white] svg{M86.3,186.2H70.9V79.1h15.4v48.4V186.2z}
                 svg{M108.9,79.1h41.6c39.6,0,57,28.3,57,53.6c0,27.5-21.5,53.6-56.8,53.6h-41.8V79.1z M124.3,172.4h24.5c34.9,0,42.9-26.5,42.9-39.7c0-21.5-13.7-39.7-43.7-39.7h-23.7V172.4z}
                 svg{M88.7,56.8c0,5.5-4.5,10.1-10.1,10.1c-5.6,0-10.1-4.6-10.1-10.1c0-5.6,4.5-10.1,10.1-10.1C84.2,46.7,88.7,51.3,88.7,56.8z};
  }
}

\newcommand\orcidicon[1]{\href{https://orcid.org/#1}{\mbox{\scalerel*{
\begin{tikzpicture}[yscale=-1,transform shape]
\pic{orcidlogo};
\end{tikzpicture}
}{|}}}}

\usepackage{kotex}

\usepackage[short,nodayofweek,level,24hr]{datetime}

\usepackage{algorithm}
\usepackage[noend]{algpseudocode}

\usepackage{pifont}

\usepackage{xcolor,color,soul}

\usepackage{hyperref}

\hypersetup{
  colorlinks   = false, 
  urlcolor     = blue, 
  linkcolor    = black,  
  citecolor    = black  
}

\usepackage{listings}

\usepackage{booktabs}
\usepackage{multirow}

\begin{document}


\title{LightSys: Lightweight and Efficient CI System for Improving Integration Speed of Software}



\author{

\IEEEauthorblockN{
Geunsik Lim$^{\textsuperscript{\orcidicon{0000-0003-1845-7132}}}$, 
MyungJoo Ham$^{\textsuperscript{\orcidicon{0000-0002-9731-0253}}}$, 
Jijoong Moon$^{\textsuperscript{\orcidicon{0000-0003-0888-2143}}}$, 
and Wook Song$^{\textsuperscript{\orcidicon{0000-0002-1642-1534}}}$
}

\IEEEauthorblockA{\textit{Samsung Research, Samsung Electronics} \\
Seoul, Republic of Korea \\
\{geunsik.lim, myungjoo.ham, jijoong.moon, wook16.song\}@samsung.com}

\thanks{Corresponding author: MyungJoo Ham, myungjoo.ham@samsung.com}

}

\maketitle

\thispagestyle{empty}%


\begin{abstract}


The complexity and size increase of software has extended the delay for developers as they wait for code analysis and code merge. With the larger and more complex software, more developers nowadays are developing software with large source code repositories. The tendency for software platforms to immediately update software packages with feature updates and bug-fixes is a significant obstacle. Continuous integration systems may help prevent software flaws during the active development of software packages, even when they are deployed and updated frequently. Herein, we present a portable and modular code review automation system that inspects incoming code changes such as code format and style, performance regression, static analysis, build and deployment tests, and dynamic analysis before merging and changing code. The proposed mechanisms are sufficiently lightweight to be hosted on a regular desktop computer even for numerous developers. The resulting reduced costs allow developers to apply the proposed mechanism to many source code repositories. Experimental results demonstrate that the proposed mechanism drastically reduces overheads and improves usability compared with conventional mechanisms: execution time (6x faster), CPU usage (40\% lower), memory consumption (1/180), and no out-of-memory occurrence.

\end{abstract}


\section{Introduction}\label{S_introduction}

Many code collaboration tools \cite{dabbish2012social, gitlab, gerrit} have been introduced to manage merging codes, including new code changes to the shared source code repository with code reviews. GitHub~\cite{dabbish2012social} is a popular collaborative code hosting platform that has been adopted by numerous commercial and open-source projects, including the Apache Software Foundation \cite{apache-github-migration}. Developers submit their source code using a pull request (PR) \cite{Nagappan} in GitHub. Developers can improve code quality by communicating with their peers in the form of code reviews before merging code changes, and using PRs will reduce defects and improve code quality.

Code review activities~\cite{sadowski-icse-seip2018, czerwonka-icse2015, Baum:2016:FIC:2950290.2950323} are required for proper software development and have certain time constraints, i.e., a code review should be published before merging or abandoning the changes. Unfortunately, many developers still regard code review as an extra task or a low-priority task, although the code review culture is crucial~\cite{sadowski-icse-seip2018}. Moreover, numerous low-quality code changes force reviewers to waste much  time and effort to point out trivial problems rather than providing more important feedback to improve the overall software project, which, in turn, discourages developers from reviewing codes. Imminent release tends to pressure developers into investing less time and effort for code review while more code changes are introduced to their projects, which significantly affects the overall project quality.

The quick introduction of many code changes significantly increases the workload on the build and integration infrastructure~\cite{tf-issue-36527}, which, in turn, may increase the time to conduct a clean build with the given changes in the infrastructure and may incur an extended delay. A clean build generates a software package without its previously built binaries and with the latest binaries of other packages in the given platform, so that the changes of other packages are fully included and accounted for. Manufacturers often increase the number of build servers by investing more resources to follow up with the increased number of build operations and their workload.

The proposed system, \textbf{\textit{LightSys}}, automates trivial parts (e.g., code format and style, defects, and bugs) of code reviews so that reviewers can focus on more important aspects of code changes. LightSys proposes novel techniques to efficiently execute various automated code review modules for incoming PRs representing instances of code changes with a highly extensible modular system and effective on-the-shelf modules. When a developer submits a PR, two groups of code review modules (pre-build and post-build) are executed. Pre-build group checks do not require building the given code and may reduce build overheads by stopping the process if the system is configured accordingly, which is the default behavior. For example, with default modules and configuration \textit{LightSys} does not build the code or test with built binaries (post-build) if the given coding convention (pre-build) is not met. Post-build group checks may rely on the build results, e.g., unit testing or platform integration and compatibility testing.

\textit{LightSys} offers enhanced building processes that allow incoming PRs to repeatedly create software packages for potential errors. The various code review and test modules enable automatic verification of the quality of code changes in the incoming PRs, where the maintainers of the corresponding projects determine the standard of code quality. Moreover, \textit{LightSys} dramatically reduces the overhead of automated code review and the continuous integration (CI) process and accelerates the overall process compared with the prior popular mechanisms. \textit{LightSys} makes the following major contributions:

\begin{itemize}
 \item A lightweight and efficient code review automation system that lowers the operating costs and accelerates the code inspection processes.
 \item A set of code inspection modules allowing reviewers to focus on more critical aspects by ensuring that the given code changes do not have minor errors, such as style errors, build errors or warnings, and failed unit test cases.
 \item A plug-in structure that is highly extensible and configurable with the concept of a plug-in store with confidence levels.
\end{itemize}

The remainder of this paper is structured as follows. Section~\ref{S_observation} describes an observation of a code review culture. Section~\ref{S_design} addresses the design and implementation of the proposed techniques in detail. Section~\ref{S_evaluation} describes our experiments and discusses their results. 
Related work is described in Section~\ref{S_related_work}. Lastly, Section~\ref{S_conclusion} concludes the paper.

\section{Observation}\label{S_observation}

\begin{figure}
\centering
\includegraphics[width=0.88\columnwidth]{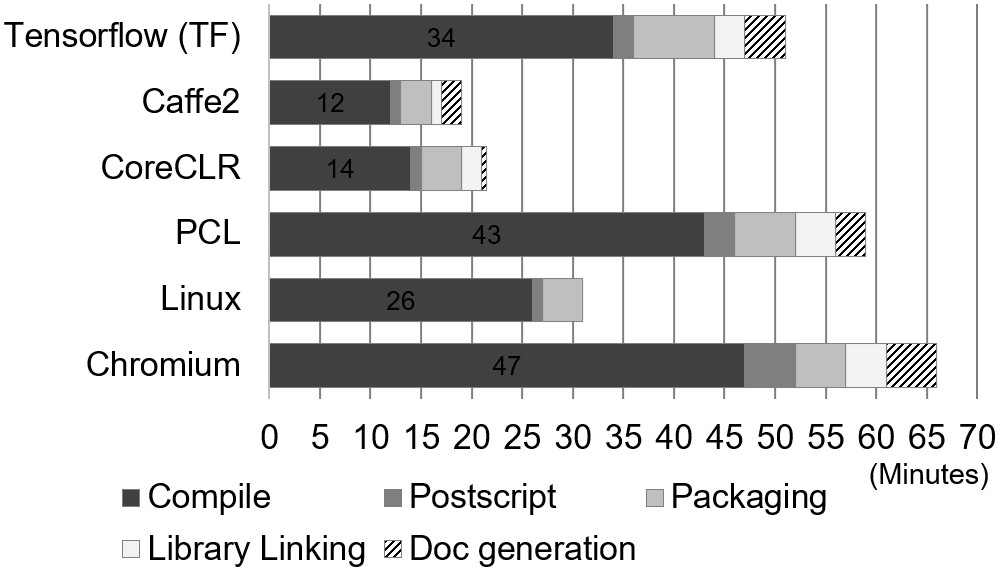}
\caption{Breakdown of build operation on massive open-source software}
\label{fig:build_breakdown}
\end{figure}

The present inefficient verification systems for code change causes delays in the deployment of  updates due to the continuous introduction of regressions and bugs and inefficient code review processes incurred by enforcing reviewers to focus on trivial aspects or discouraging review itself. As the software complexity continues to grow, frequent code changes by developers increase the risk of vulnerability to defects, regressions, bugs, and security. As a result, frequent code changes make maintaining high-quality code difficult. First, we investigated the CI usages of large open-source projects actively used in numerous deployed products. We observed the code review culture to see how embedded software developers in industry manage code quality.

\subsection{CI Usages of Open-Source Projects}\label{S_observation_pull_request}

Many large and popular open-source communities with numerous contributors have been using code review and CI services with Git. 


\textbf{Delayed Build Time}. Software is becoming more sophisticated with the growing size of each package and inter-package dependencies, which is attributable to the increased workload of build and integration \cite{8115619}. We have analyzed the build processes of six large open-source software packages, which may account for more than 50\% of the total build workload of a software platform, e.g., Tizen \cite{jaygarl2014professional, tizen-iot}. Fig. \ref{fig:build_breakdown} shows the latency breakdowns of building software in a typical desktop PC (i.e., Intel i5 CPU, 16GB DDR3, and 256GB SSD) during a CI build test. From our analysis, the compilation (``\textit{Compile}'') consumed more than 60\% of the time.

Even open-source communities with many contributors and users have been only automating build checks. Code reviews in such communities mostly rely on volunteering and most of the reviewers' time and effort is misdirected into finding trivial code flaws or styles rather than the more important aspects such as the overall structure, performance considerations, algorithms, and architectures. Surprisingly, PRs in \textit{Tensorflow} often wait for 3 to 16 days without proper attention from the CI system since it does not perform PR inspections without the reviewers’ PR approval due to the load problems of the CI server and infrastructure costs~\cite{tf-issue-36527}. These inappropriate delays are incurred by the inefficient and incorrect behaviors of CI systems~\cite{tf-issue-27939,keras-issue-12417,coreclr-issue-23265}, which often make reviewers ignore corresponding PRs and unnecessarily delay merging appropriate PRs. Such delays usually happen with large and complex projects, which incurs a huge CI workload for building the codes. Any repeated build attempts, usually triggered by PR submissions with changes of poor code quality and the build fails, further deteriorate the CI delays. As shown in Fig. \ref{fig:build_breakdown}, Open-source projects such as \textit{Tensorflow}, \textit{Point Cloud Library (PCL)}, and \textit{Chromium} require about 50 minutes or longer to build. If CI services are unstable or overloaded~\cite{caffe2-issue-17794, tf-issue-27939}, reviewers are often tempted to merge changes without trusting the CI results. Moreover, in such a situation, reviewers may simply assume that the changes are fine and approvable without actually testing them, such as reported in~\cite{pcl-issue-2924}. In the case mentioned above, the merged mainline code has actually been broken with compiler errors.

\begin{figure*}
\centering
\includegraphics[width=1.00\textwidth,height=3.9in]{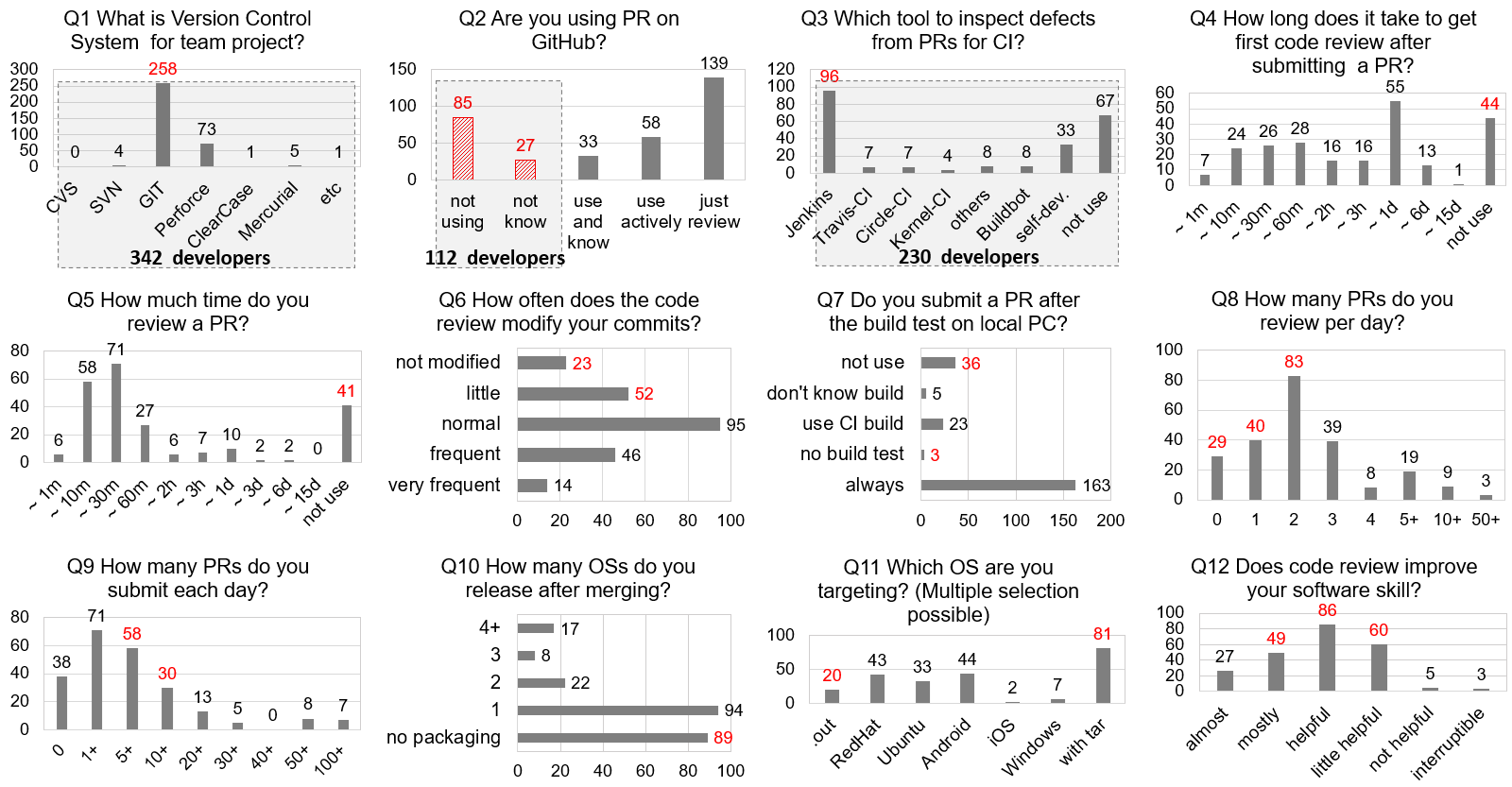}
\caption[]{Code review culture of developers in the industry. Q1--Q2 were surveyed by the 342 developers that use a version control system for their team projects. Q3--Q12 were surveyed by the 230 developers that use PRs of GitHub.}

\label{fig:observ-code-review-industry}
\end{figure*}

\textbf{Human-driven Review Process}. Our survey revealed that many open-source projects do not merge code changes without reviews to avoid regressions and bugs. The code review process with feedback via PRs (GitHub) has gained high popularity by satisfying the needs to review and merge incoming code changes made by external developers, especially with large-scale open-source projects, where contributors are supposed to have a high diversity of affiliations and locations~\cite{sadowski-icse-seip2018}. However, maintainers often merge code changes without proper review because of the lack of required check facilities or the high costs and overheads of CI services~\cite{tf-issue-26467, tf-issue-36527, coreclr-issue-23265, pcl-issue-2924}.

Lack of time or the mental cost to review code changes of other developers has been a significant reason for not using PR systems properly or in entirely. Many maintainers and reviewers are tempted to conduct only superficial code reviews. Notably, developers with no experience in code frequently review merge code changes only after negligible and superficial reviews, even though they enforce the PR procedure~\cite{10.1007/978-3-319-39225-7_8}. Merging without CI services  deteriorates the software quality significantly and a lack of CI services results in the unproductive code reviews~\cite{tf-issue-36527, tf-issue-26467, tf-issue-27939}.

\subsection{Code Review}\label{S_observation_code_review}

We have examined the code review culture that directly affects the development speed and the code quality. We have surveyed the code review community of the software developers that have deployed commercial devices in the embedded device industry. As code reviews are becoming essential to improve software quality~\cite{Baum:2016:FIC:2950290.2950323}, we have carefully examined how source codes are reviewed in collaborative developments of commercial embedded devices in the industry. As shown in Fig. \ref{fig:observ-code-review-industry}, 342 experienced developers of an internal embedded developer community of the authors’ affiliation participated in the survey research of the code review culture. They have the following experiences: a) 4 years or more experience in embedded software development, b) participated in mass software projects of consumer electronics or embedded devices, c) experience of software platforms with a feature-rich OS (Linux and Windows, not RTOS or on-chip firmware), and d) participated in team projects using GitHub.

\textbf{Passive Code Review}. Software developers in the industry are usually assigned to develop specific functions explicitly. However, code reviews often depend on volunteers and their knowledge, willingness, and spare time, and it is sometimes difficult to account for one’s own contribution and performance. Therefore, without consistent management effort and the proliferation of a proper developmental culture, it is difficult to sufficiently raise the frequency of appropriate code reviews, which can significantly improve the overall product quality and reduce developmental costs by preventing many numerous defects.

Fig. \ref{fig:observ-code-review-industry} shows how experienced developers of the embedded device industry conduct code reviews. Fig. ~\ref{fig:observ-code-review-industry}--(Q1) implies that most developers use version control systems, despite luxury that may be neglected for software in the embedded device industry. In some cases, as only one or two developers develop a software package, no proper code reviews can be expected, especially many remaining cases where code reviews are absent or not properly conducted. Code reviews are still considered a luxury if there is a strong hierarchical relation between the two developers. Automated code reviews as provided by \textit{LightSys} might significantly help such projects.

Fig. \ref{fig:observ-code-review-industry} also presents our survey of the code review culture of embedded software developers to identify the obstacles to proper code reviews. Surprisingly, the 112 developers did not use the PR process for their development projects, as shown in Q2 (\textit{not using： 85, not known： 27}). The 230 developers did not actively use a code review system such as the GitHub PR, as shown in Q4 (\textit{not use 44}) and Q5 (\textit{not use: 41}). The survey results of Q10 (\textit{no packaging： 89}) and Q11 (\textit{.out: 40, with tar: 81}) illustrate that many developers do not handle the software packaging for the OS(or software platform) deployment. Especially, Q6 (\textit{little: 52, not modified: 23}), Q7 (\textit{not use: 36, no build test: 3}), Q8 (\textit{no reviewPR: 29, review 1 PR: 40, review 2 PRs: 83}), Q9 (\textit{5-9 PRs per day: 58,10-19 PRs per day: 30}), and Q12 (\textit{helpful: 86, mostly: 49, little helpful: 60}) explain why automated code reviews may help significantly improve the code quality and PR review process.

These observations provide further motivation for our \textit{LightSys} proposal because it may alleviate these significant aforementioned issues. Notably, an automated inspection system minimizes problems, such as repetitive code defects and erroneous coding styles, and a well-designed CI system helps stabilize the entire release processes.

\section{Design and Implementation}\label{S_design}

In this section, we depict the design and implementation of our proposed system in detail to solve the non-productive situations that reviewers should repeatedly check if the source code is well written or not. The proposed system performs the pre-build checker (before the build) and the post-build checker (after the build) by separating the inspection procedure of the source code. We suggest a lightweight and efficient system that is reassembled with many user-defined check modules to improve the integration speed of source code. Our proposed system, \textit{LightSys}, automatically checks internal defects (e.g., code format, code style, static analysis, dynamic analysis, and build test) in the software and reports the results to the developers.

\begin{figure}
\centering
\includegraphics[width=0.99\columnwidth]{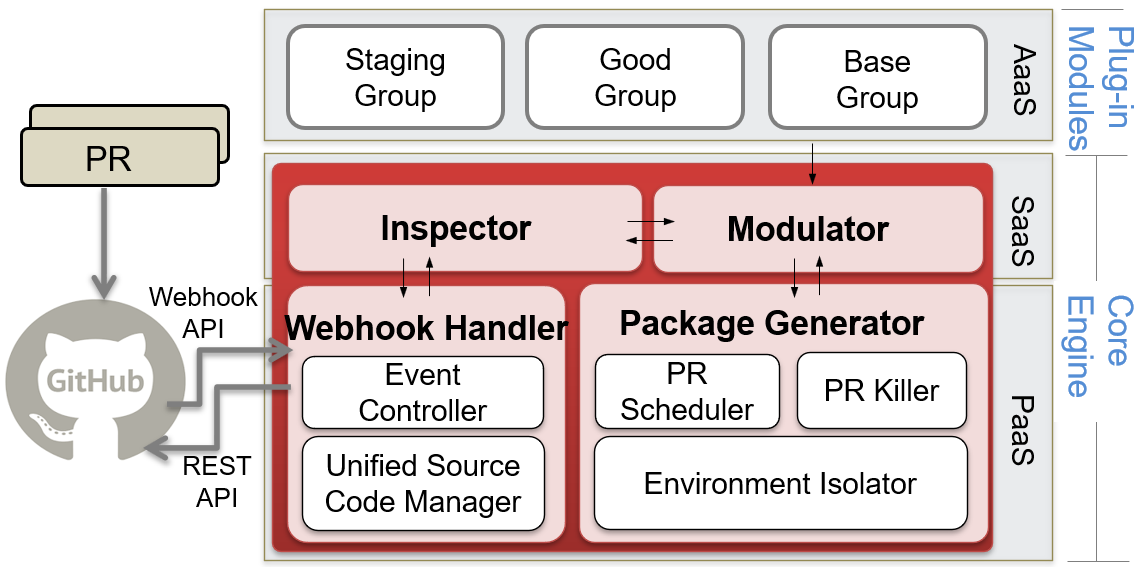}
\caption{The system architecture of \textit{LightSys}}
\label{fig:system_architecture}
\end{figure}

Fig. \ref{fig:system_architecture} shows the overall flow of \textit{LightSys}. It consists of two blocks to control a source code repository: (1) \textit{Core Engine} to conduct CI processes and (2) \textit{Plug-in Modules} to maintain three groups, namely the \textit{base} group for mandatory modules, the \textit{good} group for stable modules, and the \textit{staging} group for experimental modules, as an Application as a Service (AaaS). The AaaS service consists of modules in the form of plug-ins that can be added or removed by the user~\cite{6224294}. \textit{Core Engine} enables key components to inspect the source code submitted by developers. It consists of two services: a Platform as a Service (PaaS) such as \textit{Webhook Handler} and \textit{Package Generator}, and a System as a Service (SaaS) such as \textit{Inspector} and \textit{Modulator}.

\subsection{Webhook Handler}\label{SS_webhook_handler}

Most code hosting platforms, as well as GitHub, provide webhook API as a standard interface for managing the PR events. A \textit{Webhook Handler} consists of two components to distribute the tasks after parsing incoming event messages as follows:

\textbf{Event Controller}. Whenever a developer submits a PR to the source code repository, the repository sends a webhook event message to the specified CI server. The \textit{Event Controller} interprets the received event message and performs operations to handle the event. The \textit{Event Controller} asynchronously distributes tasks  to the \textit{Modulator} after classifying the groups of the \textit{Inspector}. Whenever developers submit their PR, the \textit{Event Controller} allocates tasks to plug-in modules via the \textit{Modulator}. In this case, the CI administrator enables or disables the tasks in the configuration file of the \textit{Event Controller}.

\begin{figure}
\centering
\includegraphics[width=0.99\columnwidth]{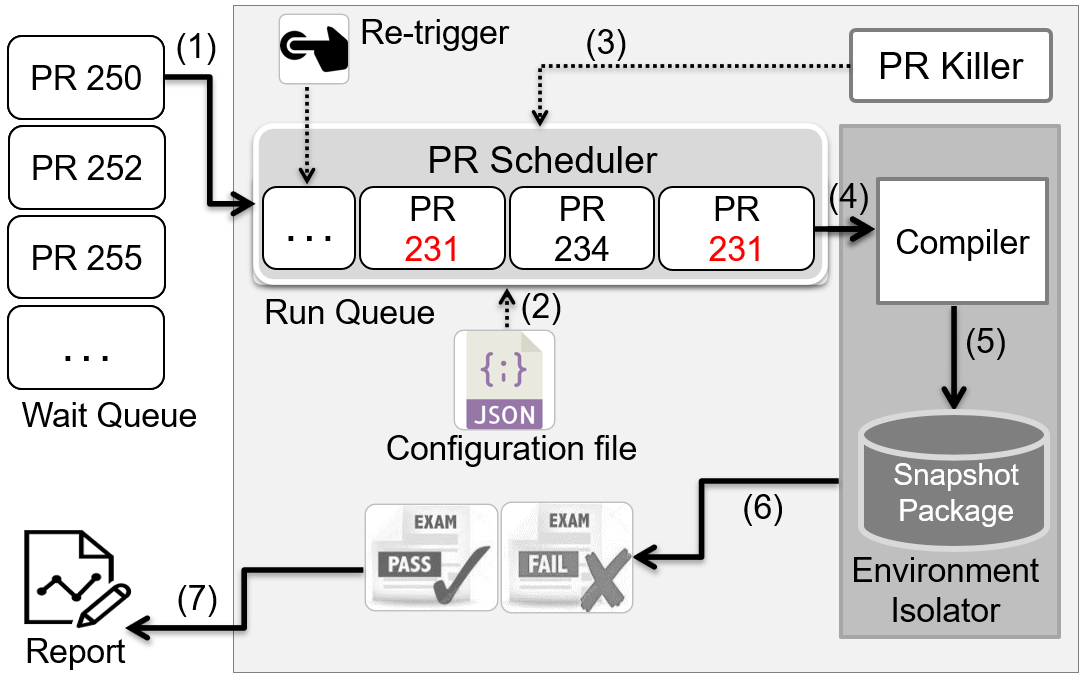}
\caption{The overall operation flow of \textit{Package Generator} to support various platforms. The numbers refer to the operation sequence of the components.}
\label{fig:package-generator}
\end{figure}

\textbf{Unified Source Code Manager}. When a developer submits a PR, the CI modules that check for the defects of the source code should use as few hardware resources as possible. In general, each check module performs a different inspection task on the PR. However, most of the modules perform a common mission of initially downloading code from the source code repository to check the source code \cite{book-humble_farley}. This task not only induces unnecessary system load on the CPU and memory of the CI server, but also creates duplicated source files in the CI server’s storage. In particular, a large quantity of source codes can overload the storage space and the maximum number of usable~\textit{inodes} of the file system~\cite{mathur:ext4}.

The \textit{Unified Source Code Manager} downloads the source code from the source code repository when a developer submits a PR. All inspection modules are then run by checking the source code by referencing the relevant source. When the test module needs to execute a test that requires modification of the source code, it performs only \textit{git branch} and \textit{git pull} operations with a \textit{Copy-On-Write (COW)} mechanism. This method dramatically minimizes the I/O operations of disk caused by additional cloning operations. These operations are performed on the same partition of the same file system so as to not generate unnecessary operations on \textit{superblock} and \textit{block bitmap} of the file system.

\subsection{Package Generator}\label{SS_package_generator}

The source code must be merged into a buildable state. The proposed system selectively supports the package build of software platforms such as \textit{Ubuntu} \cite{thomas2006beginning}, \textit{Yocto} \cite{salvador2014embedded}, \textit{Tizen} \cite{jaygarl2014professional}, and \textit{Android} for testing a clean build because the code change may affect dependent packages. It is activated when developers write a packaging script for their software platform. It provides scalability for the platform build by allowing developers to add their build modules if they need to append another software platform.

Fig. \ref{fig:package-generator} shows how \textit{Package Generator} executes the scheduling of the submitted PRs. It consists of an \textit{Environment Isolator}, \textit{PR Scheduler}, and \textit{PR Killer}. At first, developers submit their code commits with a PR to a source code repository. The \textit{PR Scheduler} is responsible for handling the requested PRs to build packages for various platforms and manages the lifespan and scheduling priority of PRs with a run queue. To deal with duplicate PRs, we designed a state-of-the-art \textit{PR Killer} that prevents unnecessary hardware resource waste.

\begin{figure}
\centering
\includegraphics[width=0.95\columnwidth]{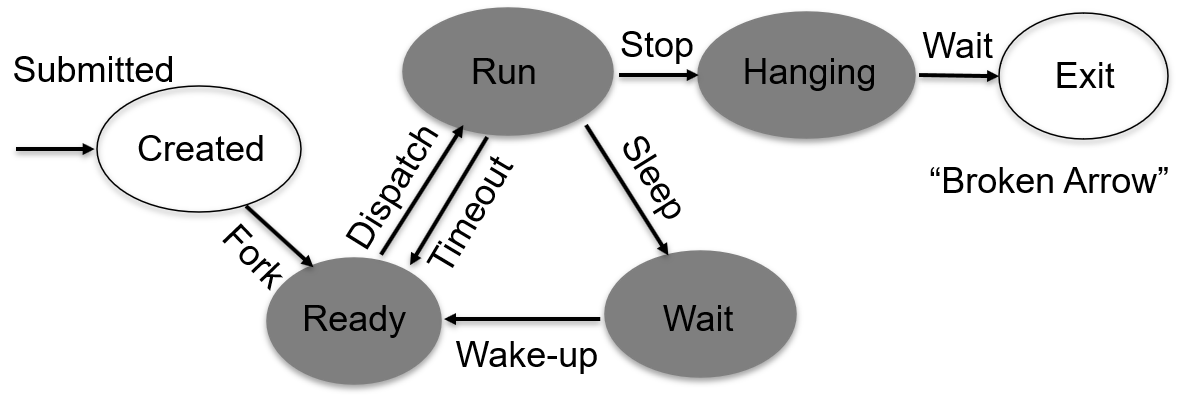}
\caption{State transition flow of PR tasks in the \textit{PR Scheduler}}
\label{fig:pr-state-transition}
\end{figure}

\textbf{Environment Isolator}. Using a virtualization technique to simultaneously build two or more source codes in an isolated environment consumes great hardware resources. Therefore, our proposed system recognizes a folder as a root folder designated to be able to build multiple source codes separately. The \textit{Environment Isolator} prepares the dependent packages and combines the source code into the packaging to establish the structure for building and testing. More than 90\% of the embedded devices are ARM CPU devices rather than X86 CPUs. Smart devices such as smartphones, smart DTVs, smartwatches, and smart refrigerators mostly use ARM CPUs. Whenever a developer submits a PR, it needs to be run on a real ARM CPU. The proposed system is designed to execute heterogeneous binary files based on \textit{QEMU-based user-space emulation} and \textit{chroot-based isolation} \cite{costin} to run the ARM CPU-based binary files on an X86 server.

\textbf{PR Scheduler}. Fig. \ref{fig:pr-state-transition} shows the overall flow of the state transition of PR. To prevent an Out-of-Memory (OOM) situation of PRs, we manage the state transition of PRs while running check modules in two groups: \textit{pre-build} and \textit{post-build}.

\begin{itemize}
\item \textit{Ready state:} The created task is ready and transited to a \textit{run state}. The \textit{PR Scheduler} selects a top-most PR according to its FIFO scheduling policy and switches to the \textit{run state}.
\item \textit{Run state:} The \textit{PR Scheduler} processes the PR when the created task is executable. The PR Scheduler executes the enabled build modules by importing the configuration file based on \textit{JavaScript Object Notation (JSON)} (See Fig. \ref{fig:package-generator}) to execute the task that enters the run state.
\item \textit{Wait state:} Many PRs are waiting to run tasks or waiting to use system resources that are locked, which means they are in sleep mode when transitioning to a \textit{wait state}.
\item \textit{Hanging state:} This state returns all system resources in use by the task to the \textit{PR Scheduler}, and the PR status is changed to the \textit{exit state}. Finally, the \textit{PR Scheduler} terminates all PIDs of the PR.
\end{itemize}

The \textit{PR Scheduler} uses two queues, namely a wait queue and a run queue, to control merge requests of source code. First, the wait queue manages an execution sequence of the created PR tasks based on First-In-First-Out (FIFO). When a job needs to be replaced in front of the wait queue, it is selected for removal. Next, the run queue contains priority values for each merge request, which will be used by the \textit{PR Scheduler}. The administrator has to configure an appropriate amount of a maximum run queue. When the job is successfully finished, the first request jobs waiting in the wait queue are moved to the run queue in sequence.

\textbf{PR Killer}. The PR often waits to receive a build ticket to perform a build operation because the \textit{PR Scheduler} does not handle a new PR when the run queue is full. After a developer submits a PR, he often resubmits the same PR (See PR 231 in Fig. \ref{fig:package-generator}) after fixing the commits due to a newly discovered defect. Moreover, developers usually modify existing PRs to apply feedback from reviewers before the previous PR build tests are completed. The \textit{PR Killer} minimizes a server overload due to duplicated PR submission. It computes the most recently submitted PR as the most stabilized source code and provides a mechanism to remove the inspection tasks of previously submitted PRs. At this time, the \textit{PR Killer} automatically kills the oldest jobs from the \textit{run state}. It is responsible for finding out-of-date tasks that must be killed based on the Least Recently Used (LRU) algorithm.

\subsection{Inspector}\label{SS_inspector}

The source code files should be independent of the CPU architecture. The source code before compilation should be able to be checked by the \textit{Inspector}. When the developer submits their PR to the source code repository, the \textit{Inspector} checks the source code level and checks for binary files that have been compiled. By dividing the work into two steps before and after compilation, \textit{LightSys} prevents unnecessary resource waste of the CI server~\cite{4599493}. The \textit{Inspector} controls the inspection procedure of PRs with separated two check structures. It consists of two components the \textit{pre-build group} and the \textit{post-build group}. When the author or reviewers close a PR, the \textit{Inspector} instantly executes an activity to destroy the running status of the existing PRs.

Most of the source code flaws can be found before the build is done. Building the source code to identify the drawbacks of software incurs a high CPU and memory cost. Therefore, our proposed system performs static code analysis and convention inspection of the source code (Pre-build group) by default, immediately informs users of an execution result when a source code defect is found, and determines whether or not to perform the next step, the build process (Post-build group), based on the available system resource of the server. That is, if a PR passes all the inspection items of the \textit{pre-build group}, the build process is always performed as a necessary procedure. Moreover, if the static code analysis needs to be run on the basis of binary codes or Java bytecode, it can be easily transferred to the \textit{post-build group} thanks to the \textit{Modulator} (described in the next section).

\begin{itemize}
\item \textbf{Pre-build group}: Run before the \textit{Package Generator} begins to compile the source. If the source code does not pass the \textit{pre-build} step, the \textit{Inspector} stops the tasks without running the \textit{Package Generator}. Also, the developer gets a report on the errors in the source code.
\item \textbf{Post-build group}: When all of the inspection modules of the \textit{pre-build group} are successfully completed for the source code, the source codes enter the post-build phase. The \textit{post-build group} validates the binary code generated after the source code is built.
\end{itemize}

\begin{figure}
\centering
\includegraphics[width=0.99\columnwidth]{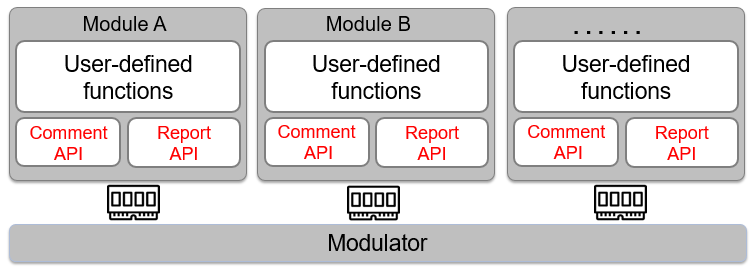}
\caption{The operation structure of the \textit{Modulator}}
\label{fig:inspector-structure}
\end{figure}


\subsection{Modulator}

The \textit{Modulator} supports a plug-in structure that allows developers to add and delete extensible check modules. At this time, the plug-in modules are classified into the following three types and operate sequentially.

\begin{enumerate}
\item \textbf{Plugins-base}: Base modules that are maintained via plug-in and plug-out, but which are always required to be performed, belong to this group.
\item \textbf{Plugins-good}: Validated source code, well-defined functions, and modules that have passed high stability tests are located in this group.
\item \textbf{Plugins-staging}: Modules that do not have enough stability and functionality to complete but have new features are in this group. Modules can be moved into the \textit{plugins-good} group when the review process and the aging test are completed. The aging test means that a module passes all test requirements without any errors for a specified period.
\end{enumerate}

Fig. \ref{fig:inspector-structure} shows how the \textit{Modulator} of our proposed system supports a customized plug-in module for source code repositories that have a different goal. CI module users can easily enable or disable custom modules with well-defined two APIs based on the plug-in structure: \textit{Comment API} sends a requested message to the GitHub repository and \textit{Report API} updates the current status of the module. At this time, the plug-in modules can be written by \textit{BASH} script in order to support easy-to-use syntax, portability, compatibility, and flexibility. The \textit{Modulator} manages three different groups based on the maturity and stability of the modules developed by the developers. In this case, the maturity is estimated by the review score of developers and the stability is measured by the aging test. The \textit{Modulator} determines where the newly added module should be executed.

\section{Evaluation}\label{S_evaluation}

\begin{table}
\small 
\caption{The execution goal of each module}
\begin{minipage}{\columnwidth}
\begin{center}
\begin{tabular}{lll}
\toprule
\textbf{No} & \textbf{Module}     & \textbf{Description}       \\
\midrule
1  &Clang-format& keep a coding standard for C/C++.       \\
2  &Cppcheck    & detect static analysis for C/C++.       \\
3  &Pylint      & check a coding standard for Python.      \\
4  &Indent      & check a code formatting style.          \\
5  &Doc-tag     & check if Doxygen syntaxes are correct.  \\
6  &Doc-build   & check if Doxygen tags are generated.    \\
7  &Scancode    & check copyrights in source code. \\
8  &File size   & check a file size to prevent too big files.\\
9  &Newline     & check if a new line exists. \\
10 &Nobody      & check if commit message is written.     \\
11 &Signed-off  & clean a license issue with Signed-off.  \\
12 &Hard-coded  & check not-allowed hard-coded paths.     \\
13 &Executable  & check if there are invalid executables. \\
14 &Timestamp   & check the timestamp of source code.   \\
15 &Sloccount   & check physical source lines of code.    \\
16 &Flawfinder  & check potential security problems in code.\\
17 &Tizen       & build a package for Tizen platform.     \\
18 &Android     & build a package for Android platform.   \\
19 &Ubuntu      & build a package for Ubuntu OS.          \\
20 &Yocto       & build a package for Yocto platform.     \\
\bottomrule
\end{tabular}
\end{center}
\bigskip
    \footnotesize
    \begin{itemize}
     \item[*] The numbers 1--7 and 15--20 are check modules based on the open source software. The \textit{pre-build group} is 1--16 and the \textit{post-build group} is 17--20.
    \end{itemize} 
\label{table:check_modules}
\end{minipage}
\end{table}


\begin{figure*}
\centering
\includegraphics[width=2.0\columnwidth]{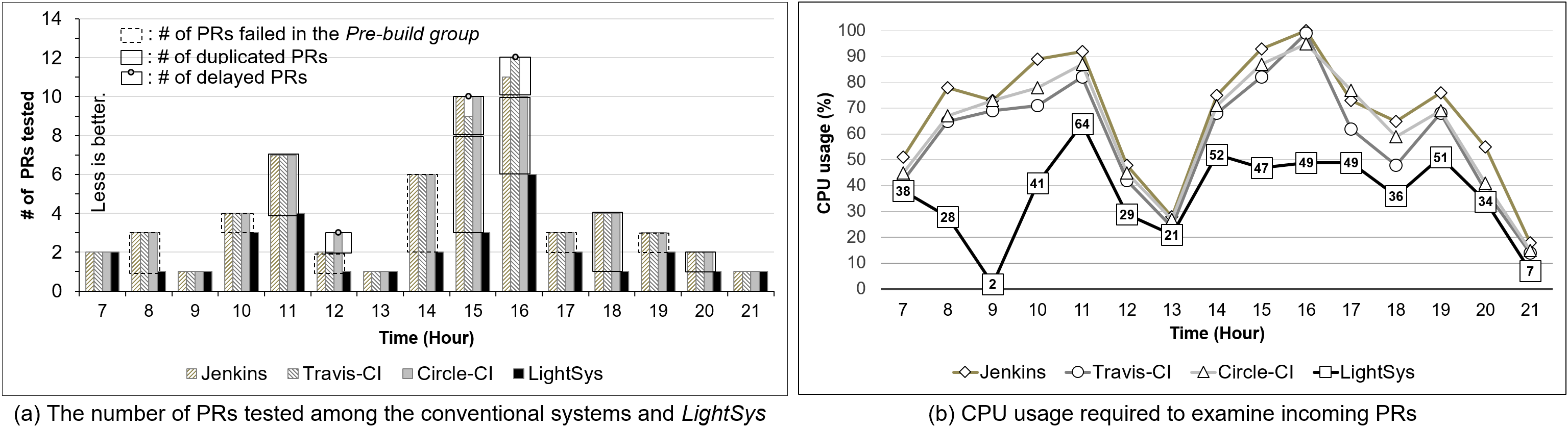}
\caption{The number of tested PRs (left) and CPU usage (right) to handle incoming PRs among three conventional systems and \textit{LightSys} from 07:00 to 21:00}
\label{fig:pr_cpu_usage}
\end{figure*}

\begin{figure}
\centering
\includegraphics[width=0.99\columnwidth]{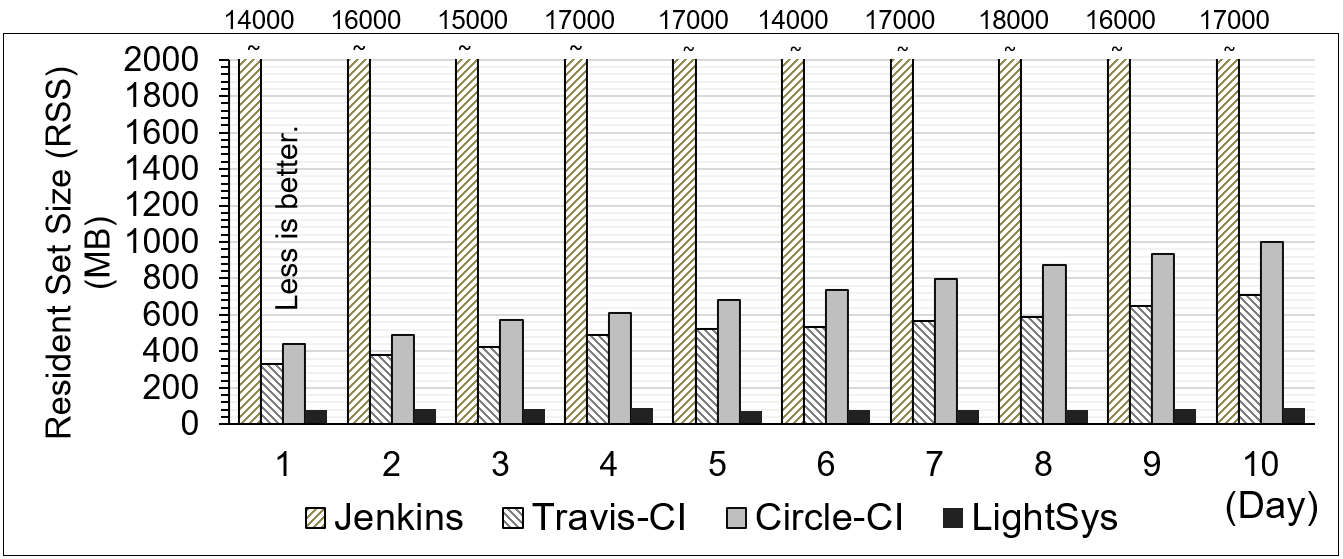}
\caption{Memory consumption comparison among three conventional systems and \textit{LightSys}}
\label{fig:memory_usage}
\end{figure}

In this section, we experimentally evaluate our \textit{LightSys} implementation. We have experimented with a public cloud occurrence of a normal desktop computer specification to show how our proposed system can run in both typical personal computers and high-performance computers. The results of the experiment are presented in response to the following four questions: (1) How does \textit{LightSys} improve the management of the check modules? (Section \ref{SS_module_enabling}), (2) How does \textit{LightSys} reduce the system resource waste? (Section \ref{eval_resource}), (3) What are the performance gains of \textit{LightSys}? (Section \ref{SS_eval_oom}), and (4) How does \textit{LightSys} improve the code integration speed? (Section \ref{eval_module_speed})

\subsection{Evaluation Setup}\label{SS_experimental_env}
The testbed is established with an AWS instance that consists of \textit{m4.4xlarge} type with Intel Xeon E5-2676 v3 2.4GHz CPU, 16 vCPUs, 16 GB Memory with Samsung DDR3, 512GB AWS-SSD/gp2 with Samsung SSD 850 PRO, 4 GB SWAP space, and Dedicated EBS Bandwidth (2000 Mbps). The software stack consists of Ubuntu 18.04 LTS and Linux kernel 4.4.15. We created three repositories (namely, \textit{Linux}, \textit{Tensorflow}, and \textit{Caffe2}) with the GitHub service for the PR-based code review process. We installed \textit{Jenkins} (an existing popular system as baseline), \textit{Travis-CI}, \textit{Circle-CI}, and \textit{LightSys} to verify the effectiveness of the proposed system. \textit{Jenkins}, \cite{smart2011jenkins} an open-source automation server, has been used in many software development projects. Although commercial software such as \textit{Travis-CI} and \textit{Circle-CI} supports professional technical support, most open-source communities still prefer to use \textit{Jenkins}, because the commercial software is not free (e.g., \textit{Travis-CI Small Business} at \$249 per a month and \textit{Circle-CI Startup} at \$129 per a month). These results arise due to the ability of Jenkins to provide users with many plug-ins, flexible plug-in extensibility, and technical mailing lists.

\begin{figure}
\centering
\includegraphics[width=0.99\columnwidth]{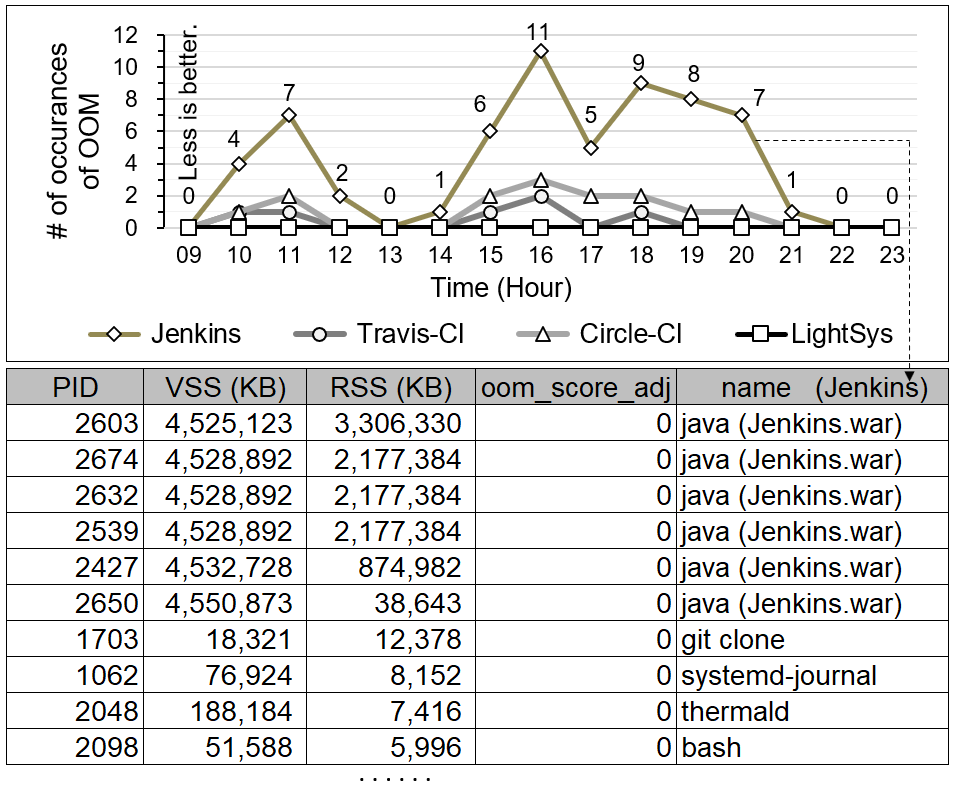}
\caption{The OOM frequency comparison among \textit{Jenkins}, \textit{Travis-CI}, \textit{Circle-CI}, and \textit{LightSys}. The table shows the 10 processes that are most likely to be killed in a descending order in the \textit{Jenkins} system.}
\label{fig:eval-oom-frequency}
\end{figure}

\begin{figure*}
\centering
\includegraphics[width=2.0\columnwidth]{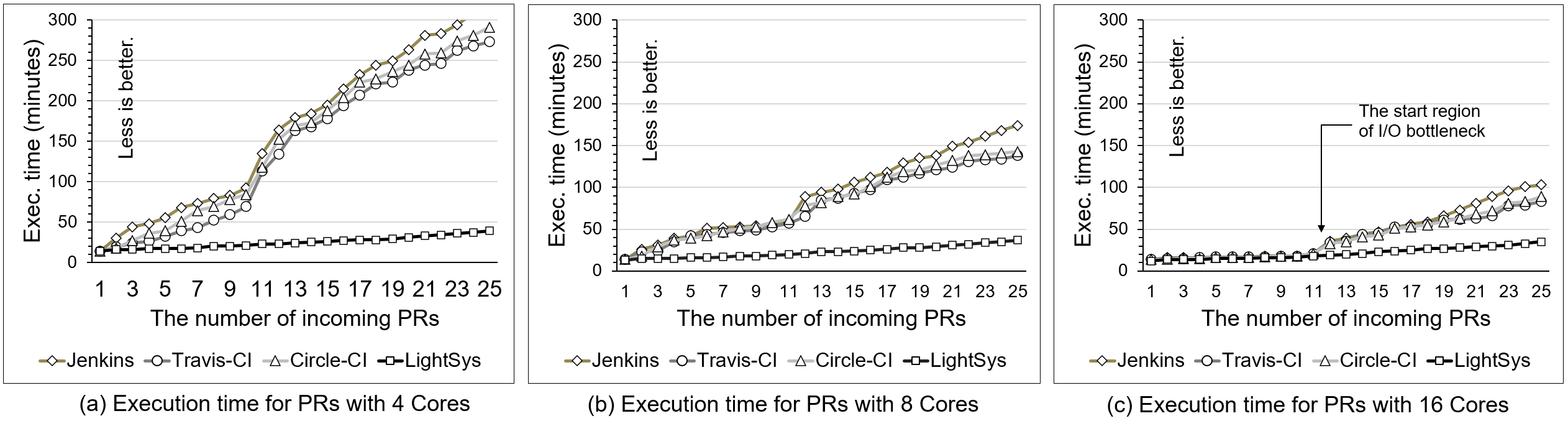}
\caption{The execution time of CI modules with multi-cores when the number of incoming PRs are increased}
\label{fig:execution_time_of_plug-in_modules}
\end{figure*}

\subsection{Enabling and Disabling Modules}\label{SS_module_enabling}

Our proposed system provides a flexible operating structure with the \textit{Modulator} compared with the existing conventional systems. As a test scenario, we randomly activated the 20 existing modules described in Table \ref{table:check_modules}. In order to establish fair comparison conditions, we executed the same \textit{BASH} scripts files among the tree conventional systems and \textit{LightSys} system. After carefully examining the tools that developers commonly use, and the activities that reviewers frequently review, we carefully developed the 20 modules. We executed an addition (\textit{+number}) and removal (\textit{-number}) of modules in consecutive order in \textit{Jenkins}, \textit{Travis-CI}, and \textit{Circle-CI}. We tested the plug-in and plug-out facilities by setting the configuration file to add and remove the modules in sequence in \textit{LightSys}. From our experiment, the conventional systems do not provide any safety equipment \cite{7965323} in case that developers have to disable or enable their inspection modules. As a result, it frequently generates code errors, duplicated codes, delicate handling of the modules due to no configuration setting and an unnecessary maintenance cost of the source code. However, \textit{LightSys} supports a configuration file to avoid the situation that an administrator has to know the internal operations of the check modules as follows:

\begin{itemize}
\item \textit{Code errors:} When the inspection modules include software bugs, developers cannot easily determine why the modules are broken because they are maintained with a Java-based plug-in structure. Such plug-in modules are mostly not intuitive as the developers have to understand Java libraries. However, \textit{LightSys} provides an easy debugging and module management mechanism with the script-based plug-in structure.
\item \textit{Code duplication:} \textit{Jenkins} has duplicated code operations among the different modules because it does not have a unified code management framework such as the \textit{unified source code manager} and \textit{Modulator}.
\item \textit{Module manipulation:} \textit{Jenkins}, \textit{Travis-CI}, and \textit{Circle-CI} does not have any mechanisms for allowing developers to manipulate the operation structure of the modules. However, \textit{LightSys} supports easy-to-use configuration files that consist of environment variables to execute internal functions selectively without modifying the modules as needed.
\item \textit{Module management:} The inspection modules of the conventional systems often generate unexpected errors when the modules are updated with new features. From our analysis, it results from the instability of the modules. Also, they do not provide a mechanism that diagnoses the execution status of the modules. However, \textit{LightSys} provides a reliable module management mechanism thanks to the well-defined \textit{Modulator} and plug-in store design.
\end{itemize}

\subsection{System Resource Usage}\label{eval_resource}

This section assesses how well the proposed system optimizes the use of system resources. Low system resource consumption not only lowers infrastructure costs but also significantly reduces the execution time of the CI modules.

\textbf{CPU Usage}. Since \textit{LightSys} inspects code changes before actually building them as explained in Section \ref{SS_inspector}, our proposed system avoids inefficient use of CPU resources. Fig. \ref{fig:pr_cpu_usage}--(a) illustrates the amount of PRs, PR frequency, and the number of PRs processed with the same number (duplicated PRs) on incoming PRs on a half-hour basis while running the CI servers from 07:00 to 21:00. Fig. \ref{fig:pr_cpu_usage}--(b) shows that the CPU resource is inefficiently wasted due to the absence of \textit{PR Killer} and \textit{Inspector}. \textit{LightSys} has reduced the CPU overhead by over 40\% by completely removing duplicated or unnecessary tasks. From our experiment, the CPU usage for inspecting incoming code changes is significantly reduced compared with that of the prior conventional systems because the state-of-the-art \textit{PR Killer} filters out PRs failed in the \textit{pre-build group} and duplicated PRs with the same number. More specifically, the prior systems process a code change even if the change is duplicated or invalidated by another.

\textbf{Memory Usage}. Fig. \ref{fig:memory_usage} shows that the memory consumption is significantly reduced with the proposed system compared with the conventional systems. In case of \textit{Jenkins}, it has Java-based VM system architecture to focus on server-based scalable software architecture. The Java virtual machine-based execution structure drastically increases memory usage. Our analysis result shows that the existing system, \textit{Jenkins}~\cite{smart2011jenkins}, is executing too many functions, even though they are not actually used. Fig. \ref{fig:memory_usage} shows how the server’s RSS memory usage can be improved with our proposed system. The RSS denotes a resident set size as the non-swapped physical memory a task has used. The y-axis shows RSS to evaluate memory efficiency experiments while running the server for 10 days without shutting down. We compared memory usage of the server in run-time without rebooting the server for 10 days for the aging test. Our experimental result showed that the proposed techniques drastically improve the frequent page reclamation and OOM~\cite{linux-oom} situation by using the minimum required memory space of our proposed system (82 MB), compared with that of the conventional systems: \textit{Jenkins} (16 GB), \textit{Travis-CI} (518 MB), and \textit{Circle-CI} (713 MB).

\subsection{OOM Frequency}\label{SS_eval_oom}

Fig. \ref{fig:eval-oom-frequency} shows our evaluation result from experimentally checking whether the proposed system enables PRs to be normally performed without OOM status, compared with existing systems in a desktop PC environment with 16 GB of physical memory. \textit{VSS} means a virtual set size as the total amount of virtual memory used by the task. \textit{RSS} means a resident set size, which is the non-swapped physical memory used by a task. \textit{oom\_score\_adj} is a heuristic value in order to adjust the final memory score of the OOM killer in out-of-memory conditions. As expected, the experimental results show that the proposed system does not have any OOM operation (x- axis: 0 times) thanks to the \textit{PR Scheduler} and \textit{PR Killer} when executing the check modules for the PR. Existing systems generate frequent OOMs. When the conventional systems process more PRs than it can handle, the Linux kernel runs the OOM killer to obtain available memory space for the new PR. When developers submit many PRs, the OOM frequency is higher. The OOM situation occurs because  the conventional systems do not have a PR management mechanism such as the \textit{PR scheduler} of the proposed system. As a result, the Linux kernel forcibly terminates the existing PRs, at which time the conventional systems cannot switch the PR state because it does not know of this termination \cite{jenkins-oom-2012, jenkins-oom-2015, jenkins-oom-2019}. As a result, the running (or waiting) state of PR \cite{tf-issue-36527, keras-issue-12417, coreclr-issue-23265} is more than 15 days. The table of Fig. \ref{fig:eval-oom-frequency} shows that the Linux kernel selects victim processes according to the \textit{RSS} and \textit{oom\_score\_adj} \cite{linux-oom}. As a result, the CI execution state of the PR is maintained in the running state for several days.


\subsection{Execution speed of plug-in modules}\label{eval_module_speed}

Fig. \ref{fig:execution_time_of_plug-in_modules} compares the execution latency of CI modules when we increase incoming PRs, with the 20 plug-in modules described in Table \ref{table:check_modules}. We experimented the build latency of \textit{Caffe2}, one of the popular deep learning frameworks with four build modules (See No. 17--20 explained in Table \ref{table:check_modules}). From our analysis, the conventional systems generate additional build costs because they compile source codes with a Docker container \cite{thalheim2018cntr, sochat2018} that consists of \textit{aufs}, \textit{process namespace}, and \textit{cgroup}. However, the \textit{Lightsys} provides users with a faster build process thanks to the lightweight \textit{Environment Isolator} depicted in Section \ref{SS_package_generator}. The conventional systems duplicates many tasks because there is no unit to manage the integrated modules. Many operations are repeated among the modules because the existing system does not unify most of the duplicated processes (e.g., repeated git clone, unnecessary functions, and absence of the \textit{PR Killer}). It is complicated to work with standard and shared procedures due to the existing naive design structure. Moreover, the execution speed of the conventional systems at compile time is directly dependent on the performance of the multi-core hardware. In the conventional systems, the execution speed of the entire modules can be improved when it allocates as many CPUs as there are modules, while executing this procedure in parallel. However, this increase in the number of incoming PRs degrades the I/O performance, so that the execution speed of the build modules gradually increases the latency. From our analysis, the I/O bottleneck arose when the number of incoming PRs reached 11. As a result, we could determine that a naive system design tackles the high performance of Solid-State Drive (SSD) storage. The proposed system provides the \textit{Modulator} that consistently controls the functions to share and use modules commonly. The \textit{Modulator} thereby helps developers develop new modules by simplifying the implementations of new functionalities, as shown described in Fig. \ref{fig:inspector-structure}. Therefore, the developers can focus on the development of the features.


\section{Related Work}\label{S_related_work}

We summarize prior work on CI, defect inspection, PR-based code reviews, and code reviews in general. PR-based code reviews, along with CI services, are popular for many open-source projects while we still have many large and favorite projects not using pull-requests (or similar methods): GCC and Glibc.

\textbf{CI/CD Tools on GitHub}. \textit{Syed} \cite{syed2018jenkins} presents an approach to set up a CI technique in a project using Jenkins and Artifactory. However, Jenkins does not provide developers with a client-based CI script management scheme compared with \textit{Travis CI}.  \textit{Zhao et al.}~\cite{8115619} suggest a more nuanced picture of how GitHub teams are adapting to and benefiting from CI services. The empirical study presents the impact of adopting \textit{Travis CI}~\cite{6802994} on a collection of GitHub~\cite{Vasilescu} projects, and survey the developers who adopted \textit{Travis CI}. \textit{Containershare} \cite{sochat2018} presents an open-source library of containers, both providing itself as a template, a library, and a production API. This paper describes the building and testing of source code with a container-based CI tool, \textit{Circle-CI}. However, \textit{Jenkins}, \textit{Travis-CI}, and \textit{Circle-CI} do not handle the performance issues of large projects with excessively numerous code changes. Moreover, since \textit{Travis-CI} and \textit{Circle-CI} are commercial software, many software developers prefer to use open-source software, \textit{Jenkins}. 
\textit{TAOS-CI} \cite{icce-taosci} proposes a modulable mechanism that inspects source codes with many user-defined check modules to reduce the review time of developers on incoming patches into the GitHub development repository. However, these papers do not provide a mechanism to deal with duplicate PRs, a configurable plug-in structure for, and a fine-grained light-weight check technique for efficient system overheads.

\textbf{Inspecting Defects of PR}. \textit{Pham et al.}~\cite{6606557} present insights in the contribution process of GitHub~\cite{Nagappan} and show how maintainers inspect code changes. Their research identifies challenges and solutions in practice; however, their solutions solely rely on the discretion of maintainers and reviewers without any assistance from automated systems for the code quality. \textit{VCCFinder}~\cite{ccs2015-vccfinder} proposes a method to find potentially dangerous code with a significantly lower false-positive rate than comparable systems. Their method combines code-metric analysis with meta-data gathered from code using machine-learning techniques. Their weakness is that vulnerabilities of initial commits may persist on the deployed software.

\textbf{Code Review in General}. \textit{Modern Code Review}~\cite{sadowski-icse-seip2018} investigates motivations and practices of code reviews at Google and the developers’ satisfaction and challenges. They find that code reviews are important to the workflow at Google. The high cost and overestimated benefits of code reviews have often discouraged code reviews. \textit{Jacek}~\cite{czerwonka-icse2015} emphasizes the need for more sophisticated guidelines of code review workflows. Code refactoring is a common practice for improving maintainability and readability. \textit{Why We Refactor?}~\cite{silva-fse2016} finds that code refactoring is usually triggered by the introduction of new features and bugs. They propose that refactoring recommendations should refocus from code-smell-oriented to maintenance-task-oriented solutions.

\section{Conclusion}\label{S_conclusion}

This paper proposes lightweight automated code checking mechanisms that are highly effective in achieving rapid and low-cost CI-assisted effective code reviews, which, in turn, help to generate higher code quality with fewer delays. The proposed system also includes a novel mechanism enabling the developers to employ various configurable code inspectors for many popular Git-repository managers (e.g., GitHub, GitLab, and Gogs). The maintainers can quickly develop or extend inspection modules with the given plug-in structure. Moreover, this paper introduces state-of-the-art code change inspection techniques that swiftly and lightly re-assemble and manage source codes. \textit{LightSys} is available as fully open-source software at \cite{github-taosci} and has been applied to many commercial software projects for embedded systems in the affiliation since it enables developers to reduce non-productive code review time and improve integration speed.


\section*{Acknowledgment}

We would like to thank the anonymous reviewers, Sewon Oh, Parichay Kapoor, Dongjoo Chae, Sangjung Woo, Minho Ju, Inki Dae, Gwanghoon Son, Hong-Seok Kim, Duil Kim, and Daehyun Kim, for their insightful comments and feedback that helped improve the paper. This work was supported in part by On-Device AI Computing (RAJ0121ZZ-35RF), Samsung Research, Samsung Electronics Co., Ltd.


\normalsize 

\bibliographystyle{IEEEtran} 


\bibliography{ref}

\end{document}